# Eco-Friendly Supercapacitor Architecture Based on Cotton Textile Waste and Biopolymer-Based Electrodes

Luis T. Quispe[a,*], Clemente Luyo Caycho[b], Javier Quino-Favero[c], Silvia Ponce[d], Abel Gutarra[a,*]

[a] Grupo de Investigación en Economía Circular, Instituto de Investigación Científica (IDIC), Universidad de Lima, Av. Javier Prado Este 4600, Fundo Monterrico Chico, Surco, Lima 15023, Perú
[b] Center for the Development of Advanced Materials and Nanotechnology, National University of Engineering, Av. Tupac Amaru 210, Lima 15333, Perú
[c] Carrera de Ingeniería Ambiental, Grupo de Investigación en Soluciones Tecnológicas para el Medio Ambiente, Universidad de Lima, Av. Javier Prado Este 4600, Fundo Monterrico Chico, Surco, Lima 15023, Perú
[d] Carrera de Ingeniería Industrial, Grupo de Investigación en Economía Circular, Instituto de Investigación Científica (IDIC), Universidad de Lima, Av. Javier Prado Este 4600, Fundo Monterrico Chico, Surco, Lima 15023, Perú

**Abstract**

This study presents an eco-friendly and bio-based symmetric supercapacitor using cotton textile waste-derived hydrogels as electrolytes and chitosan-based carbon electrodes as metal-free charge-storage components. Cotton-derived hydrogels were synthesized via an alkaline dissolution–gelation route and modified with ammonium thiocyanate ($NH_4SCN$) to enhance ionic conductivity. The ionic modification increased the hydrogel conductivity from 17.1 to 37.8 mS cm$^{-1}$, confirming a nearly twofold improvement in ion transport efficiency. The resulting hydrogel exhibited improved thermal stability, homogeneous ionic transport, and strong polymer-ion interactions confirmed by FTIR and TGA analyses. In a symmetric device, the ion-modified hydrogel enables reduced equivalent series resistance, faster charge-transfer kinetics, and a short time constant ($\tau = 3.2$ s), comparable to commercial energy-storage systems. After 1000 cycles, the device exhibits a 12.3 % increase in specific capacitance, confirming stable proof-of-concept operation. Cycling leads to a moderate increase in $R_{ESR}$ (18 → 22 Ω) and $\tau$ (3.2 → 4.1 s), indicating slower charge-ion redistribution. Notably, this $R_{ESR}$ includes the contribution of the test-cell setup; in compact coin-type configurations, the resistance would be considerably lower. EIS reveals a concurrent rise in interfacial resistive terms, consistent with post-cycling hydrogel darkening and FTIR evidence of Fe-SCN coordination, suggesting that resistance buildup mainly originates from minor Fe-SCN interactions when the expelled liquid reaches the stainless-steel collector, rather than from loss of capacitive functionality. Overall, these results demonstrate the viability of cotton waste-derived hydrogels and chitosan-based electrodes as sustainable components for green energy storage, offering a recyclable and eco-friendly alternative to conventional systems.

**Keywords:** Eco-friendly supercapacitor; cotton textile waste; cotton hydrogel electrolyte; chitosan-based electrodes; ionic modification; green energy storage.

# 1. Introduction

The global shift toward renewable and sustainable energy systems has intensified the demand for eco-friendly and bio-based materials capable of replacing petroleum-derived components in energy storage devices [1,2]. Among these devices, electrochemical supercapacitors (SCs) have emerged as promising candidates due to their high power density, long cycle life, and rapid charge-discharge capabilities [3–6]. However, conventional supercapacitor architectures often rely on synthetic polymer binders and non-renewable electrolytes, which generate environmental concerns associated with toxicity, carbon footprint, and end-of-life disposal [7–9]. To address these issues, research has increasingly focused on bio-based, recyclable, and low-impact materials that enable high electrochemical performance without compromising environmental sustainability [10,11].

Natural polymers such as cellulose, chitosan, and alginate have gained considerable attention as sustainable building blocks for energy devices owing to their biodegradability, hydrophilicity, and ability to support efficient ionic conduction [12,13]. In particular, cellulose-based hydrogels possess a unique three-dimensional network structure capable of retaining large amounts of water and ions, thus serving as solid or quasi-solid electrolytes in green supercapacitors [14–16]. Moreover, cellulose can be sourced from post-consumer cotton waste, which is primarily composed of pure β-(1→4)-linked D-glucopyranose units, providing both high purity and mechanical strength [17,18]. The valorization of such textile residues into functional hydrogel matrices contributes to the goals of circular economy and waste minimization [19].

Recent studies have demonstrated the feasibility of hydrogels derived from cotton or cellulose fibers for diverse applications. Azam et al. [20] prepared alginate hydrogels reinforced with cotton fibers, enhancing their tensile strength and water retention for biomedical purposes. Chumpitaz et al. [21] synthesized magnetic cotton-based hydrogels for controlled drug release, highlighting the chemical adaptability and biocompatibility of cotton-derived cellulose. Furthermore, Landi et al. [10] and Klobukoski et al. [22] showed that biopolymer-based hydrogel electrolytes can provide competitive ionic conductivity and stability when used in eco-friendly supercapacitors. These findings support the concept of using recycled cellulose waste as both structural and electrochemical material in sustainable devices.

Despite these advances, reports addressing hydrogels derived specifically from cotton textile waste for use as electrolytes in supercapacitors remain scarce [23–25]. The integration of such hydrogels with bio-based and metal-free electrodes represents a largely unexplored path toward high-performance, low-impact energy storage systems. Chitosan, a natural polysaccharide derived from chitin, has been identified as a biopolymeric binder that enhances mechanical integrity, adhesion, and proton transport while replacing toxic fluorinated polymers like PVDF [26–28]. When combined with activated carbon or carbon black, chitosan-based electrodes can deliver excellent capacitance and stability in aqueous environments [10,29].

In this work, we propose a fully sustainable symmetric supercapacitor architecture composed of (*i*) cotton textile waste-derived hydrogels as electrolytes, and (*ii*) chitosan-based carbon electrodes as eco-friendly, metal-free charge storage components. This approach aligns with the United Nations Sustainable Development Goals (SDG 12 and 13) by promoting responsible material use and reduced environmental impact in electrochemical systems [30]. The resulting device demonstrates that bio-based and eco-friendly materials can achieve electrochemical performance comparable to conventional systems while maintaining recyclability, and adherence to circular economy principles [10,11,19].

## 2. Materials and Methods

### 2.1 Sample Preparation

The SCs developed in this study are symmetric devices, composed of two identical electrodes and a hydrogel electrolyte formulated in an aqueous medium. In line with the principles of sustainable materials engineering, the hydrogel electrolytes were prepared using cotton-derived waste fibers, while the electrode inks incorporate biopolymeric binders. The preparation methods and compositions of each component are detailed in the following sections.

#### 2.1.1. Fabrication of Electrodes Using Bio-Based and Metal-Free Materials

*(i)* Preparation of the binder: The binder solution was prepared by dissolving 1.7 g of commercial chitosan (powder, deacetylation degree > 75%, viscosity 20 - 300 cP, Sigma-Aldrich, USA) in an aqueous solution of 1.0 wt% glacial acetic acid (ACS reagent grade, ≥ 99.7 %, J.T. Baker, USA), adjusting the quantities so that the final solution contained 2.43 wt% chitosan. The chitosan was dissolved under constant magnetic stirring at room temperature for approximately 12 h, until a homogeneous, slightly viscous solution free of visible lumps was obtained.

*(ii)* Preparation of the active material ink: The active material ink was prepared by mixing 2.7 g of activated charcoal "AC" (powder, DARCO® G-60, 100 mesh, specific surface area 500 - 600 $m^2$ $g^{-1}$, Sigma-Aldrich, USA), 0.3 g of carbon black "CB" (Super P Li, specific surface area ~62 $m^2$ $g^{-1}$, primary particle size ≈ 40 nm, TIMCAL, China), and 7.0 g of a chitosan-based binder solution. The mixture was subjected to alternating cycles of manual agitation and ultrasonic bath treatment until all visible agglomerates of AC and CB were eliminated, yielding a uniform and homogeneous ink.

*(iii)* Impregnation of the ink onto the current collector: A commercial carbon cloth (FuelCellStore, USA) was employed as the current collector due to its high electrical conductivity through-plane (< 13 mΩ $cm^2$) and chemicall stability in acidic environments. Prior to ink application, the carbon cloth was cleaned in an ultrasonic bath containing isopropyl alcohol for 10 minutes to remove surface contaminants and improve adhesion. The prepared active material ink was applied directly onto the surface of the cleaned carbon fabric using a soft-bristle brush, ensuring even coating across the electrode surface. The amount of ink deposited was carefully controlled so that, after drying and complete evaporation of the aqueous content, the final active material load was approximately 4 mg cm$^{-2}$, in accordance with the recommendations reported by M. D. Stoller *et al.* [1]. The pristine carbon cloth used as the current collector had an initial thickness of about 0.33 mm, which increased to approximately 0.42 mm after the deposition and drying of the active material layer.

(*iv*) Drying and finalization of the electrodes: After ink application, the samples were dried in an oven at 50 °C overnight to remove water and residual acetic acid from the binder. This step ensured good adhesion of the active material to the current collector and preserved the porous structure of the active layer. Once dried, the resulting films constituted the final electrodes.

2.1.2. Preparation of Sustainable Hydrogel Electrolytes from Cotton Waste Fibers

(*i*) Pulverization of cotton textile waste: The starting material for the hydrogel electrolyte was post-consumer textile waste composed of Pima cotton fabric (thread count 20/1). These fabric scraps were manually cut into squares of approximately 1 cm$^2$. To obtain fibrous particles, the pieces were processed in a high-speed blade mill operating at 20000 rpm. As the milling proceeded, lighter fiber fragments were lifted by the upward air stream, reaching a height of about 30 cm above the blades, and subsequently collected using a fine mesh filter. This process yielded cotton microfibers with cross-sectional diameters typically ranging from 10 to 20 μm and longitudinal dimensions extending to several millimeters.

(*ii*) Wet milling, sieving, and freeze-drying: The cotton powder obtained from the initial milling step was further processed using a planetary ball mill (model TOB-YXQM-2L, Xiamen TOB New Energy Technology Co., China) equipped with four agate vessels. In each vessel, 3.5 g of pulverized cotton was dispersed in water at a fiber-to-water mass ratio of approximately 3%. Agate milling balls were used to avoid material contamination. Milling was performed for 15 hours at 180 rpm to yield cotton microfibers suitable for subsequent hydrogel preparation. After milling, the resulting suspension was sequentially passed through sieves with mesh openings of 250 μm and 75 μm. This allowed obtaining a fraction of microfibers with a granulometry from 75 to 250 μm, which was subjected to a freeze-drying process to obtain dry cotton microfibers.

(*iii*) Hydrogel synthesis: The hydrogels were synthesized using an alkaline gelation-dissolution method using previously freeze-dried cotton microfiber fractions [21]. For each formulation, 0.5 g of microfibers was dispersed in 10 mL of an aqueous alkaline solution composed of 7 wt% sodium hydroxide (NaOH, pellets for analysis, EMSURE®, Merck, Germany) and 12 wt% urea (ACS reagent, grade for analysis, Supelco-Merck, Germany). The dispersion was transferred to cylindrical molds and subjected to freezing at -24 °C for 2 hours. Subsequently, a 5 wt% citric acid (ACS reagent, ≥ 99.5 %, Sigma-Aldrich, Canada) solution was gently applied dropwise over the frozen precursors using a pipette, ensuring complete surface coverage. Without delay, the molds were placed in an ultrasonic bath and treated for 5 minutes to accelerate the chemical crosslinking process. After crosslinking, the formed hydrogels were removed from the molds and rinsed with deionized water to remove residual reagents. Due to the alkaline composition of the precursor solution, the hydrogel inherently contains sodium ($Na^+$) and hydroxide ($OH^-$) ions in its polymer network, which contribute to the intrinsic ionic character of the hydrogel. The resulting hydrogel contains approximately 10 wt% cellulose-based solid network and 90 wt% water-rich phase confined within the polymer matrix. This hydrogel will henceforth be referred to as "H-1".

(*iv*) Ionic Modification of the Hydrogel: To enhance the electrochemical performance of the system, the H-1 hydrogel was subjected to an ionic incorporation process. Specifically, the H-1 hydrogels were immersed in an aqueous 5 wt% ammonium thiocyanate ($NH_4SCN$, crystals, ACS reagent, 99.0% purity, J.T. Baker, USA) solution and maintained under magnetic stirring at 50 °C for 1 hour. This treatment enabled the controlled diffusion of $NH_4^+$ and $SCN^-$ ions into the hydrogel network, promoting an increased density of mobile charge carriers within the polymeric matrix. Hereafter, the ion-enriched hydrogel obtained through this procedure will be referred to as "H-2".

2.1.3. Assembly of symmetrical supercapacitors:

The assembly of the supercapacitors was carried out using a sandwich-type architecture, in which the hydrogel simultaneously served as both electrolyte and separator. Two identical electrodes were placed face-to-face, with a thin layer of the corresponding hydrogel interposed between them. All the hydrogels used exhibited a disk-shaped geometry, with an average thickness of approximately 2 mm and a diameter of about 13 mm. Two types of symmetric supercapacitors were fabricated to investigate the influence of ionic incorporation on the electrochemical behavior. The configuration electrode/H-1/electrode was designated as SC-1, whereas the configuration electrode/H-2/electrode was referred to as SC-2.

Figure 1a displays the two electrodes and the hydrogel film required for the sandwich-type assembly. Figure 1b shows the assembled supercapacitor just before being placed inside the hermetic stainless-steel split test cell (TOB-3ESTC15P, SUS-304) used for electrochemical characterizations. The use of the test cell ensures a hermetic seal and uniform contact pressure between the electrodes, preventing hydrogel dehydration and guaranteeing reproducible conditions during electrochemical characterizations. The applied pressure was regulated through the displacement of the compression screw, adjusted so that the total deformation of the hydrogel did not exceed 0.1 mm. Figure 1c presents a schematic representation of the SC-1 device, highlighting the symmetric arrangement of the electrodes and the H-1 hydrogel layer positioned at the center of the assembly.

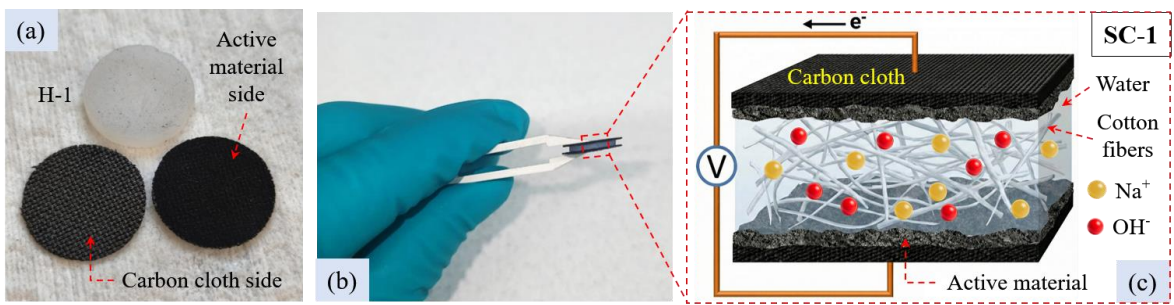

**Figure 1.** (a) Individual components used in the sandwich-type configuration, showing the two identical electrodes and the H-1 hydrogel layer. (b) Photograph of the assembled supercapacitor. (c) Schematic illustration of the SC-1 device, highlighting the symmetric architecture with the H-1 hydrogel positioned between the two carbon-based electrodes.

2.2 Characterizations

2.2.1. Thermogravimetric (TGA)

TGA analysis of the hydrogels H-1 and H-2 was conducted using a Labsys Evo 1150 (Setaram, France) thermal analyzer. Approximately 10 mg of each sample was placed in alumina crucibles and heated under a continuous nitrogen flow of 15 mL min$^{-1}$ to prevent oxidation. The measurements were performed within the temperature range of 25 - 800 °C, applying a heating rate of 10 °C min$^{-1}$. This analysis was carried out to evaluate the thermal stability and decomposition behavior of the hydrogels before and after ionic incorporation.

2.2.2. Fourier transform infrared spectroscopy (FTIR)

Lyophilized hydrogels H-1, H-2 and the hydrogel recovered from the cycled SC-2 device (H-2 after cycling) were analyzed by Fourier Transform Infrared (FTIR) spectroscopy. The measurements were carried out using a Nicolet iS10 spectrometer (Thermo Scientific, USA) equipped with an Attenuated Total Reflectance (ATR) accessory. All samples were examined directly on the ATR platform without additional preparation, employing a diamond crystal as the reflective element to ensure intimate contact between the sample surface and the detector. Spectra were collected in absorbance mode over a wavenumber range of 4000 - 650 cm$^{-1}$, using a spectral resolution of 4 cm$^{-1}$ and averaging 20 scans per sample to enhance the signal-to-noise ratio. Prior to each measurement, the ATR crystal was cleaned with ethanol and verified for baseline stability to guarantee data reliability.

2.2.3. Electrochemical characterizations

The electrochemical impedance spectroscopy (EIS) measurements were carried out to evaluate the electrochemical behavior of the individual hydrogel electrolytes and the assembled supercapacitor device. These analyses were performed using an Autolab electrochemical workstation. The frequency range was set from 100 kHz to 0.1 Hz with an AC amplitude of 10 mV at open-circuit potential. The ionic conductivity (σ, S cm$^{-1}$) of the hydrogel electrolytes was calculated using the bulk resistance (R$_b$, Ω), which was determined from the intercept of the Nyquist plot on the real axis (Z'). The calculation was performed using Eq. (1) [25],

$$\sigma = \frac{d}{R_b S} \tag{1}$$

where $d$ is the thickness of the hydrogel sample (cm) and $S$ is the contact surface area between the electrolyte and the electrode (cm$^2$). To evaluate the electrochemical performance of the supercapacitors SC-1 and SC-2, cyclic voltammetry (CV) and galvanostatic charge-discharge (GCD) tests were conducted using the same Autolab workstation. The CV curves were recorded within a potential window of 0 to 1.0 V at scan rates of 10, 20, 30, 40, and 50 mV/s. The GCD measurements were performed at current values from 3 mA (0.48 A g$^{-1}$) to 6 mA (0.96 A g$^{-1}$) for SC-1, and from 7 mA (1.12 A g$^{-1}$) to 15 mA (2.40 A g$^{-1}$) for SC-2. The cyclic stability of the SC-2 was evaluated by subjecting the device to 1000 charge-discharge cycles at a current value of 15 mA (2.40 A g g$^{-1}$) within a potential window of 0 to 1.0 V.

The specific capacitance values commonly reported in commercial supercapacitor datasheets correspond to a single electrode ($C_{sp-e}$) [1], which is also considered in this work. However, to facilitate comparison with studies that report the specific capacitance of the complete device ($C_{sp-SC}$), the relationship between both parameters for our symmetric SC is presented in Eq. (2) [1,5]. Both capacitances were calculated from the galvanostatic charge-discharge (GCD) curves.

$$C_{sp-e} = 4C_{sp-SC} = 4\frac{I}{m(\Delta V/\Delta t)} \tag{2}$$

where $I$ is the discharge current, $\Delta t$ is the discharge time (s), $\Delta V$ is the potential window (V) excluding the IR drop, $m$ is the total mass of the active material in both electrodes (g). The factor 4 is due to the symmetrical nature of the two-electrode system. Furthermore, the specific energy (E, Wh kg$^{-1}$) and specific power (P, W kg$^{-1}$) of the supercapacitors SC-1 and SC-2 were calculated using Eq. (3) and Eq. (4):

$$E = \frac{1}{2}C_{sp-SC}(\Delta V)^2 \tag{3}$$

$$P = \frac{E}{\Delta t} \tag{4}$$

## 3. Results and Discussion

3.1 Thermal Behavior of Hydrogels

Figure 2 shows the thermogravimetric (TGA) and derivative (DTG) curves of hydrogels H-1 and H-2, revealing the influence of ionic incorporation with $NH_4SCN$ on their thermal stability and degradation profile. In H-1, the first region below 150 °C corresponds to the removal of physically adsorbed and weakly hydrogen-bonded water [20]. The second region (150–350 °C) corresponds to thermal degradation of cellulose and citric acid crosslinks, with DTG peaks at 189 °C and 268 °C, associated with glycosidic bond cleavage, dehydration, and depolymerization of the cellulose backbone [31]. Above 350 °C, gradual weight loss occurs from oxidation and breakdown of carbonaceous residues, in two stages (360 - 500 °C and 560 - 780 °C). The final residue of ~10.9 wt% reflects partial carbonization of the cellulose structure [31]. H-2 follows a similar thermal profile but exhibits enhanced stability. The dehydration peak shifts slightly to ~124.6 °C, indicating stronger hydrogen bonding and ionic interactions between $NH_4^+/SCN^-$ and cellulose hydroxyl groups [20]. The main degradation peaks appear at 215 °C and 327 °C, both shifted to higher temperatures relative to H-1, with broader DTG signals reflecting a slower decomposition process stabilized by ionic crosslinking and C-N-S-type bonding [13]. Above 350 °C, the weight loss is minimal, and the residue reaches ~26.9 wt%, suggesting the formation of carbonitride- and sulfur-containing species derived from $NH_4SCN$ decomposition [32]. These results confirm that ionic modification with $NH_4SCN$ improves the structural cohesion and thermal resistance of the cellulose hydrogel. Notably, TGA studies on cotton-derived hydrogels remain limited in the literature, underscoring the relevance of these findings for understanding the thermo-structural stability of bio-based ionic networks [12].

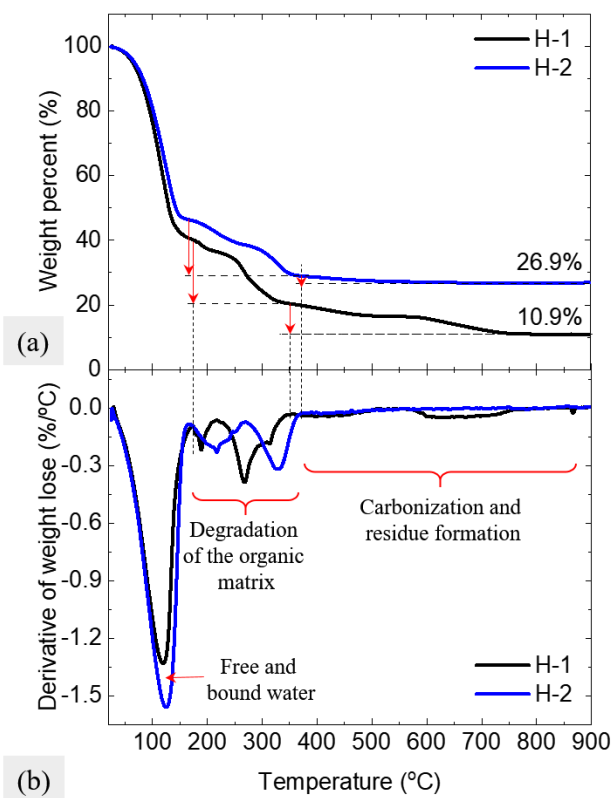

**Figure 2.** (a) TGA and (b) DTG curves of hydrogels H-1 and H-2. Both display typical multi-step weight loss behavior related to dehydration, organic matrix degradation, and carbonization. The NH$_4$SCN-modified hydrogel (H-2) shows improved thermal stability compared with the pristine sample (H-1).

3.2 FTIR analysis

Figure 3 shows the FTIR spectra of hydrogels H-1, H-2 (before cycling), as well as the hydrogel recovered from the SC-2 device after electrochemical cycling tests (H-2 after cycling). All samples display the characteristic absorption bands of the polymeric matrix and the functional groups generated upon ionic incorporation. The broad bands observed in the 3400 - 3200 cm$^{-1}$ region correspond to O–H and N–H stretching vibrations, while those at 2890 cm$^{-1}$ and 1650 - 1600 cm$^{-1}$ are associated with C-H and C=O/C-N stretching modes of the hydrogel backbone [20,21]. The characteristic spectral feature associated with thiocyanate ions appears in the 1900 - 2200 cm$^{-1}$ region. The right panel of Figure 3 magnifies this range, where the distinctive band of the thiocyanate anion (SCN$^-$) is clearly observed. In hydrogel H-2, a peak centered at approximately 2060 cm$^{-1}$ is identified and assigned to the C≡N stretching vibration of free or weakly coordinated SCN$^-$ species [12,32]. In the case of H-2 after cycling, an additional band emerges at around 2096 cm$^{-1}$, identified by Lorentzian fitting as a secondary component associated with interactions between SCN$^-$ ions and metallic species (Fe, Cr, Ni) originating from the stainless-steel components of the test cell (SUS304 steel).

Previous studies have demonstrated that thiocyanate ions can induce transient coordination with iron, promoting the formation of Fe-SCN-type species ($[Fe(SCN)]^{2+}$, $[Fe(SCN)_2]^+$) responsible for FTIR bands in the 2090 - 2100 cm$^{-1}$ range [33,34]. In aqueous and electrochemical environments, such interactions are linked to partial breakdown of the passive film and exposure of active iron sites, where SCN$^-$ acts as a coordinating or complexing ligand [34]. Therefore, the appearance of this additional band can be attributed to the formation of Fe-SCN complexes at the hydrogel-metal interface, an undesirable reaction that indicates coordination or charge-transfer processes capable of compromising the chemical integrity of the hydrogel and reducing the long-term stability of the supercapacitor.

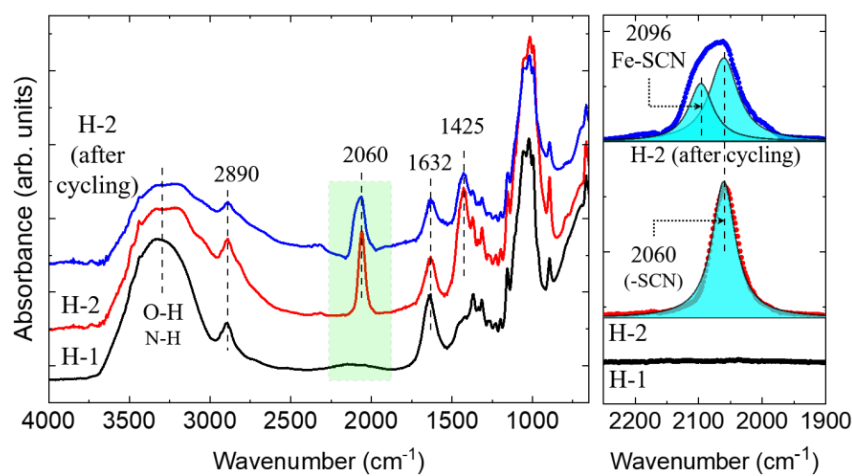

**Figure 3.** FTIR spectra of hydrogels H-1, H-2, and H-2 after cycling. The left panel shows the overall spectral range, while the right panel highlights the 1900 - 2250 cm$^{-1}$ region, where a new band at ~2096 cm$^{-1}$ appears after cycling, attributed to Fe-SCN-type interactions.

3.3 Electrochemical Characterization of the Supercapacitors

Prior to assembling the symmetric supercapacitor devices, the ionic transport properties of the individual hydrogels were evaluated by EIS, as shown in Figure 4a. The H-2 curve is clearly shifted toward lower Z′ values, evidencing a reduced total resistance and enhanced ionic transport. Figure 4b highlights the high-frequency region used to determine the bulk resistance ($R_b$), yielding $R_b$ = 6.2 Ω for H-1 and $R_b$ = 2.8 Ω for H-2. Using Eq. (1) and considering identical thickness (t = 1.7 mm) and contact area (S = 1.61 cm$^2$), the corresponding ionic conductivities were $\sigma_{(H-1)}$ = 17.1 mS cm$^{-1}$ and $\sigma_{(H-2)}$ = 37.8 mS cm$^{-1}$. These results confirm that ionic incorporation with NH$_4$SCN nearly doubles the ionic conductivity, enhancing charge mobility through the cotton hydrogel network and explaining the improved electrochemical response of devices using H-2.

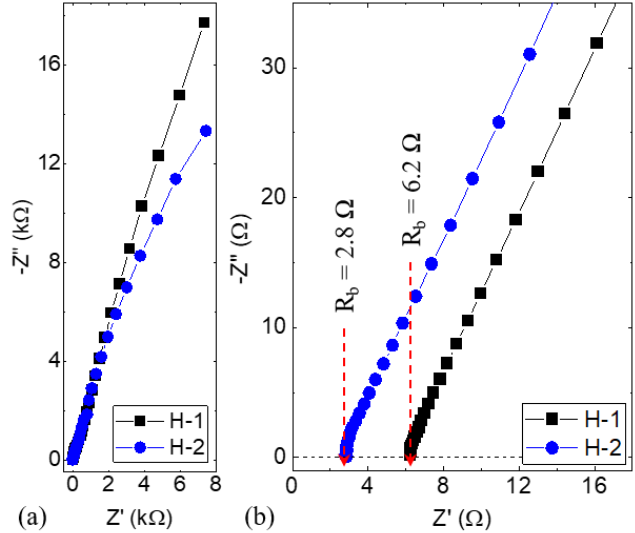

**Figure 4.** Electrochemical impedance spectroscopy (EIS) spectra of hydrogels H-1 and H-2. (a) Full Nyquist plots recorded over the 100 kHz - 0.1 Hz frequency range. (b) Magnified view of the high-frequency region near the intercept with the real axis (Z′).

Electrochemical measurements were also performed to evaluate the charge-storage performance and interfacial behavior of the symmetric supercapacitors SC-1 and SC-2. Cyclic voltammetry (CV), galvanostatic charge-discharge (GCD), and electrochemical impedance spectroscopy (EIS) were employed to analyze their capacitive response, internal resistance, and cycling stability. Figure 5 presents the cyclic voltammetry (CV) curves of the symmetric supercapacitors SC-1 and SC-2 recorded at different scan rates. Neither device displays pronounced redox peaks, confirming the predominance of electrostatic double-layer charge storage typical of carbon-based capacitors [35]. For SC-1 (Figure 5a), the CV curves retain a near-rectangular shape at low scan rates, suggesting efficient ion adsorption at the electrode-hydrogel (H-1) interface. At higher scan rates, a slight distortion and a transition toward a leaf-shaped profile are observed, indicating increased kinetic limitations associated with ion diffusion through the hydrogel matrix [35,36]. This behavior aligns with models describing the exponential decrease in capacitance with scan rate due to restricted ion motion within porous electrodes [36]. In SC-2 (Figure 5b), which employs the ion-modified hydrogel (H-2), the CV curves remain more rectangular compared to SC-1, even at elevated scan rates, suggesting improved ionic transport and charge distribution within the electrode/electrolyte interface. The incorporation of $SCN^-$ ions enhances ionic conductivity and facilitates faster charge accumulation, consistent with previous findings on gel polymer electrolytes containing mobile ionic species [13,37]. The broader current response and larger enclosed area of SC-2 compared to SC-1 confirm its superior capacitive behavior, associated with enhanced ion accessibility and lower polarization losses [35,36].

Figures 5c and 5d display the galvanostatic charge-discharge (GCD) profiles of the symmetric supercapacitors SC-1 and SC-2, respectively, at various current densities. Both devices exhibit nearly linear and symmetric triangular shapes, confirming the capacitive nature of the charge-storage mechanism, consistent with double-layer behavior and negligible faradaic contribution [22,35]. In SC-1 (Figure 5c), the discharge curves show a slight IR drop at the onset of discharge ($R_{ESR} \approx 65 - 70\ \Omega$), which reflects resistive effects related to ionic transport within the pristine hydrogel (H-1). This behavior is typical of polymeric electrolytes with moderate ionic mobility, where limited charge carrier diffusion induces voltage polarization at higher current densities [12,13]. The discharge times decrease gradually with increasing current, indicating stable but diffusion-limited charge transfer processes. In contrast, SC-2 (Figure 5d), incorporating the ion-modified hydrogel (H-2), exhibits longer discharge durations and smaller IR drops at equivalent current densities ($R_{ESR} \approx 17 - 18\ \Omega$), demonstrating enhanced ionic conductivity and reduced interfacial resistance. The presence of $SCN^-$ ions within the hydrogel network increases the number of mobile charge carriers, facilitating faster ionic migration and improved electrode/electrolyte coupling [13,32,38]. This enhancement translates into higher specific capacitance and improved energy efficiency. As will be detailed in the subsequent EIS analysis, the observed behavior is supported by a lower charge-transfer resistance and enhanced diffusive response, confirming the beneficial effect of ion incorporation in SC-2. Overall, the superior discharge behavior of SC-2 confirms the beneficial effect of ion incorporation on the electrochemical dynamics and stability of the hydrogel electrolyte. This improvement in electrochemical performance is consistent with the higher ionic conductivity observed for the H-2 hydrogel (Figure 4), which reduces resistive losses and facilitates faster ion migration across the electrode/electrolyte interface.

The specific capacitance per single electrode ($C_{sp-e}$) for both supercapacitors SC-1 and SC-2 was calculated using Equation (1), derived from the discharge section of the galvanostatic charge-discharge (GCD) curves. This approach follows the standard method recommended for evaluating electrical double-layer capacitors (EDLCs), where the slope of the discharge profile provides a direct measure of the stored charge and energy efficiency [1,39,40]. As shown in Figure 5e, both devices exhibit the expected decrease in capacitance with increasing current density, reflecting the limited ion diffusion and polarization effects at higher discharge rates. However, SC-2 consistently delivers higher $C_{sp-e}$ values across all tested current densities, confirming that the ion-modified hydrogel (H-2) enables more efficient charge accumulation. This improvement arises from enhanced ionic conductivity and better electrode-electrolyte coupling, which facilitate faster ion migration within the porous electrode matrix [11].

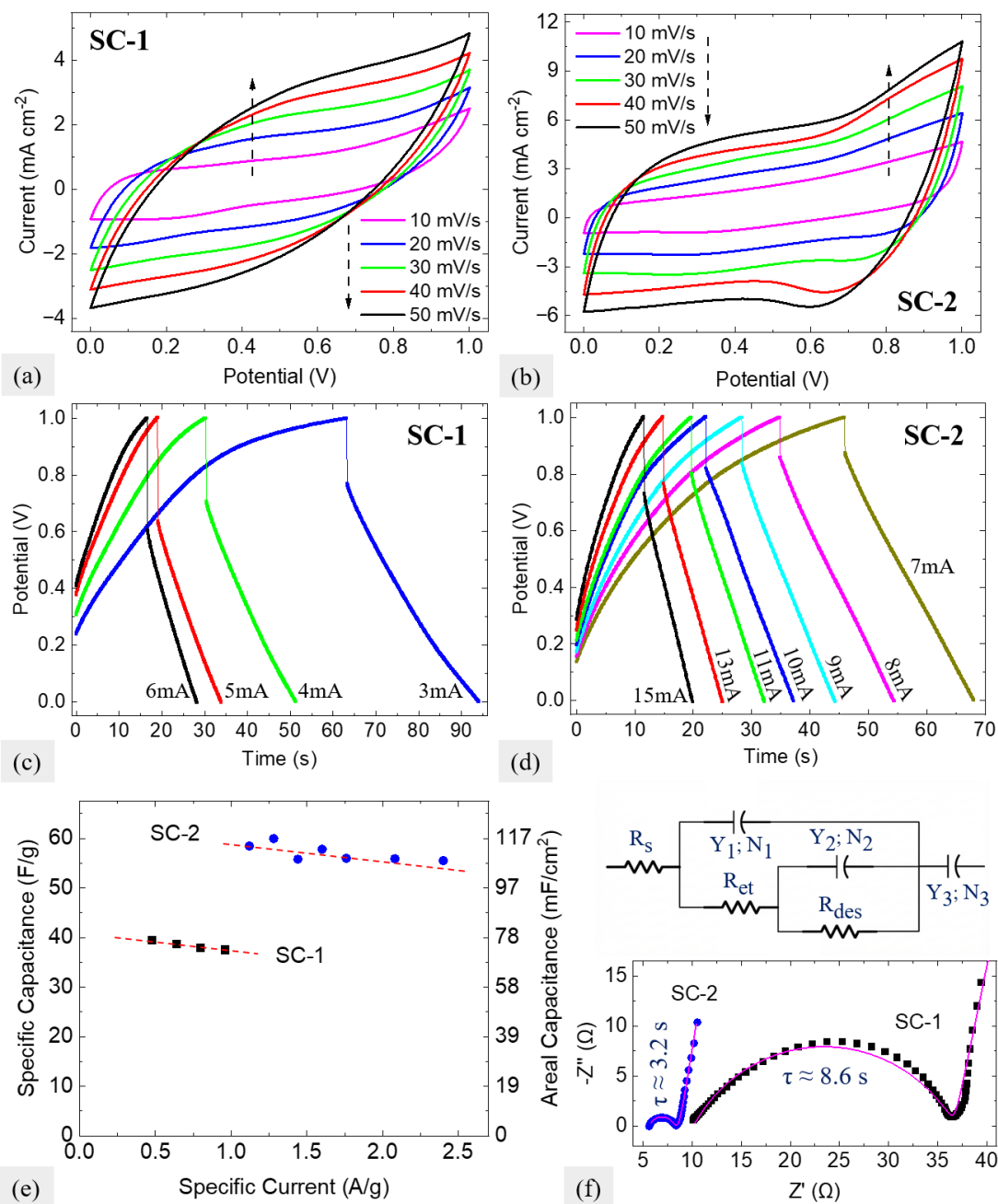

**Figure 5.** Electrochemical characterization of the symmetric supercapacitors SC-1 and SC-2: (a) CV curves of SC-1 recorded at different scan rates; (b) CV curves of SC-2 under identical conditions, showing improved capacitive response with ion-modified hydrogel. (c) GCD profiles of SC-1 at various current densities; (d) GCD profiles of SC-2, exhibiting longer discharge times and reduced potential drop. (e) Variation of specific capacitance $C_{sp-e}$ as a function of current density. (f) Nyquist plots of SC-1 and SC-2 obtained from electrochemical impedance spectroscopy (EIS), including the equivalent circuit model used for fitting.

Figure 5f shows the Nyquist plots of the symmetric supercapacitors SC-1 and SC-2, together with their fittings obtained using the equivalent circuit model depicted in the inset. The fitting results provided the electrochemical parameters that describe resistive and capacitive processes occurring in the system. The obtained values are summarized in Table 1, which lists the solution resistance ($R_s$), the charge-transfer resistance ($R_{et}$), and the ionic desorption resistance ($R_{des}$); the latter representing the resistance encountered by ions during their release from the electrode surface or pore structures back into the electrolyte [?]. Additionally, the Table 1 includes the parameters of the constant phase elements (CPEs) associated with the double-layer ($Q_{dl}$: $Y_1$, $N_1$), diffusive ($Q_d$: $Y_2$, $N_2$), and pseudocapacitive ($Q_{ps}$: $Y_3$, $N_3$) responses [41].

As anticipated from the electrolyte analysis in Figure 4, the H-2-based device (SC-2) exhibits lower resistive components at all levels of the equivalent circuit model. For SC-1, the larger intercept on the real axis ($R_s = 10.1\ \Omega$) and broader semicircle ($R_{et} = 21.9\ \Omega$) reflect limited ionic conductivity and sluggish interfacial charge transfer within the pristine hydrogel (H-1). The additional $R_{des} = 5.34\ \Omega$ suggests a slower ionic release from the electrode surface, indicating that part of the charge remains confined within microporous domains. The double-layer element ($Y_1 = 69.8\ \mu Mho \cdot s^N$, $N_1 = 0.676$) confirms a non-ideal capacitive behavior caused by surface heterogeneity, while the low exponent in the diffusive component ($N_2 = 0.180$) and the small admittance ($Y_2 = 9.0\ mMho \cdot s^N$) point to hindered ion diffusion and weak transport coupling across the hydrogel/electrode interface. The pseudocapacitive component ($Y_3 = 104\ mMho \cdot s^N$, $N_3 = 0.906$) suggests limited faradaic participation. Furthermore, using $\tau = R_{ESR} \times C_{SC}$ [5], the overall time constant for SC-1 is $\tau \approx 8.6$ s, pointing to relatively slow charge-discharge relaxation, consistent with diffusion-restricted kinetics.

In contrast, SC-2 exhibits lower $R_s$ (5.43 $\Omega$), $R_{et}$ (2.99 $\Omega$), and $R_{des}$ (0.25 $\Omega$), evidencing enhanced ionic conductivity and faster interfacial kinetics. The higher admittance values ($Y_1 = 90.0\ \mu Mho \cdot s^N$, $Y_2 = 78.5\ mMho \cdot s^N$, $Y_3 = 146\ mMho \cdot s^N$) demonstrate improved double-layer formation, efficient ion diffusion, and stronger pseudocapacitive coupling. Moreover, the near-unity exponents $N_2 = 0.961$ and $N_3 = 0.890$ indicate a highly homogeneous charge-distribution regime, with well-balanced faradaic and capacitive contributions. Importantly, the characteristic time constant for SC-2 ($\tau \approx 3.2$ s) is significantly shorter than that of SC-1, confirming a faster electrochemical response and more efficient ion transport dynamics [38,42]. Moreover, this $\tau$ value falls within the typical range reported for most commercial supercapacitors (0.5 - 3.6 s) [4,5], supporting the practical relevance of the fast response of SC-2. Overall, the combined interpretation of Figure 5f and Table 1 demonstrates that ionic modification of the hydrogel substantially decreases both resistive and kinetic barriers, resulting in a faster relaxation process and enhanced capacitive efficiency for SC-2, consistent with the preceding CV and GCD analyses.

**Table 1.** Fitting parameters obtained from the equivalent circuit model (shown in Figure 5f) applied to the EIS data of SC-1 and SC-2. Additionally, the last column shows the time constant ($\tau = R_{ESR} \times C_{SC}$) associated with the supercapacitor.

| Sample | $R_S$ Ω | $R_{et}$ Ω | $R_{des}$ Ω | $Y_1$ µMho·s$^N$ | $N_1$ - | $Y_2$ mMho·s$^N$ | $N_2$ - | $Y_3$ mMho·s$^N$ | $N_3$ - | $\tau$ s |
|---|---|---|---|---|---|---|---|---|---|---|
| SC-1 | 10.10 | 21.90 | 5.34 | 69.8 | 0.676 | 9.0 | 0.180 | 104 | 0.906 | 8.6 |
| SC-2 | 5.43 | 2.99 | 0.25 | 90.0 | 0.649 | 78.5 | 0.961 | 146 | 0.890 | 3.2 |

The electrochemical stability of the SC-2 device was assessed through galvanostatic charge-discharge (GCD) cycling over 1000 consecutive cycles, and the calculated parameters are summarized in Figure 6. As shown in Figure 6a, the specific capacitance gradually increases during the initial cycles and stabilizes thereafter, resulting in an overall capacitance enhancement of approximately 12.3%. A similar initial increase in capacitance during the first 1000 cycles has also been reported by other research groups [6]. This improvement is attributed to the activation of electroactive sites and the enhanced ionic accessibility promoted by the ion-modified hydrogel (H-2), which improves charge transfer at the electrode-electrolyte interface [38]. In Figure 6b, both the specific energy and specific power exhibit a slow decline as cycling progresses, consistent with minor internal resistance buildup. Nevertheless, the retention of high performance after 1000 cycles demonstrates the durability and resilience of the system under repeated operation.

Figure 6c shows the Nyquist plots of the SC-2 device, recorded before the stability test and after 500 and 1000 charge-discharge cycles, along with their respective fittings using the same equivalent circuit shown in Figure 5f. The obtained fitting parameters are summarized in Table 2. A progressive increase in the resistive elements is observed after cycling ($R_s$: 5.43 → 6.02 Ω; $R_{et}$: 2.99 → 6.96 Ω; $R_{des}$: 0.25 → 0.48 Ω), indicating a decline in ionic conductivity and less efficient interfacial transport. These changes arise from partial blocking of electroactive sites and increased interfacial heterogeneity, as suggested by the EIS fitting results. Concurrently, the high-frequency element $Q_{dl}$ decreases ($Y_1$: 90.0 → 49.5 µMho·s$^N$; $N_1$: 0.649 → 0.552), evidencing a loss of double-layer ideality. The mid-frequency element $Q_d$ increases ($Y_2$: 78.5 → 110.0 mMho·s$^N$) while $N_2$ slightly decreases (0.961 → 0.890), which reflects structural reorganization of the hydrogel and broadening of ionic diffusion pathways. The pseudocapacitive contribution $Q_{ps}$ remains relatively stable ($Y_3$: 146 → 140 mMho·s$^N$; $N_3$: 0.890 → 0.863), confirming that the long-timescale capacitive process remains active [41,42]. Finally, the characteristic time constant ($\tau$) increases with cycling (Table 2), indicating slower device-scale relaxation and ion-transport kinetics. This trend is consistent with progressive interfacial heterogeneity and partial blocking of electroactive sites during cycling, which delay charge redistribution.

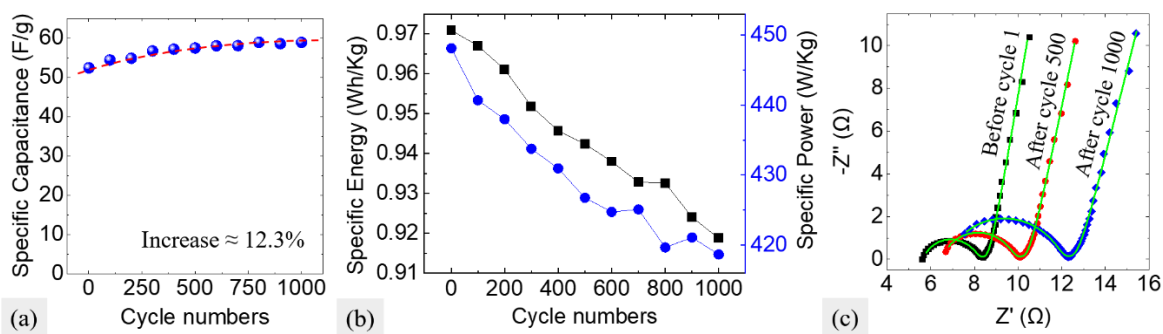

**Figure 6.** Electrochemical stability analysis of the SC-2 device after 1000 charge-discharge cycles. (a) Evolution of specific capacitance $C_{sp-e}$ as a function of cycle number, showing an overall increase of approximately 12.3% after 1000 cycles. (b) Variation of specific energy and specific power with cycling, indicating moderate losses and stable performance after 1000 cycles. (c) Nyquist plots obtained before cycling and after 500 and 1000 cycles, illustrating slight changes in charge-transfer resistance and the preservation of the capacitive response, confirming the electrochemical stability of the device SC-2.

It is important to note that, after 1000 cycles, the hydrogel showed a progressive dark brown coloration, which can be attributed to secondary interactions between $SCN^-$ ions and Fe species released from the stainless-steel (SUS304) current collector. Although the hydrogel does not directly contact the collector, the liquid fraction expelled under compression can wet the carbon cloth and reach the metallic surface, allowing limited and transient coordination between Fe and $SCN^-$ ions, leading to the formation of Fe-SCN-type complexes ($[Fe(SCN)]^{2+}$, $[Fe(SCN)_2]^+$) [33,34]. This process partially modifies the electrolyte composition and is consistent with the FTIR results (Figure 3), where the appearance of a new band at ~2096 cm$^{-1}$ confirms the presence of such Fe-SCN interactions after cycling. Therefore, the increased impedance and interfacial non-ideality observed in the EIS spectra are attributed to electrolyte alteration caused by these coordination processes rather than intrinsic electrode degradation. Despite this limitation, the SC-2 device maintains functional capacitive behavior up to 1000 cycles, validating the proof-of-concept feasibility of the ion-modified hydrogel system for sustainable energy storage applications. Despite this limitation, the SC-2 device maintains functional capacitive behavior up to 1000 cycles, validating the proof-of-concept feasibility of the ion-modified hydrogel system for sustainable energy storage applications.

**Table 2.** Fitting parameters obtained from the equivalent circuit model (shown in Figure 5f) applied to the EIS spectra of the SC-2 supercapacitor before cycling and after 500 and 1000 charge-discharge cycles. The last column shows the corresponding time constant values ($\tau = R_{ESR} \times C_{SC}$) associated with the supercapacitor.

| SC-2 | $R_S$ Ω | $R_{et}$ Ω | $R_{des}$ Ω | $Y_1$ μMho.s$^N$ | $N_1$ - | $Y_2$ mMho.s$^N$ | $N_2$ - | $Y_3$ mMho.s$^N$ | $N_3$ - | $\tau$ s |
|---|---|---|---|---|---|---|---|---|---|---|
| Before Stability Test | 5.43 | 2.99 | 0.25 | 90.0 | 0.649 | 78.5 | 0.961 | 146 | 0.890 | 3.2 |
| After Cycle 500 | 5.82 | 4.38 | 0.31 | 72.0 | 0.584 | 89.5 | 0.926 | 145 | 0.869 | 3.7 |
| After Cycle 1000 | 6.02 | 6.96 | 0.48 | 49.5 | 0.552 | 110.0 | 0.890 | 140 | 0.863 | 4.1 |

Figure 7 shows the Ragone plot of the developed supercapacitors (SC-1 and SC-2), where their specific energy and specific power values are compared with the typical ranges reported for different electrochemical energy storage systems [2]. Both devices fall within the characteristic region of symmetric supercapacitors, exhibiting a favorable balance between specific energy and specific power. A remarkable improvement in the performance of SC-2 compared to SC-1 is observed, attributed to the ionic incorporation process applied to the hydrogel, which enhances charge transport and improves the efficiency of the electrode-electrolyte interface. This modification results in higher specific energy values without compromising power delivery capability, demonstrating more efficient electrochemical kinetics in the device with the modified hydrogel. It is noteworthy that, despite being fabricated from recycled and renewable-source materials, both devices exhibit electrochemical properties comparable to those reported in the literature for conventional symmetric supercapacitors, validating the feasibility and potential of the eco-friendly materials employed in this work for the development of sustainable energy storage technologies.

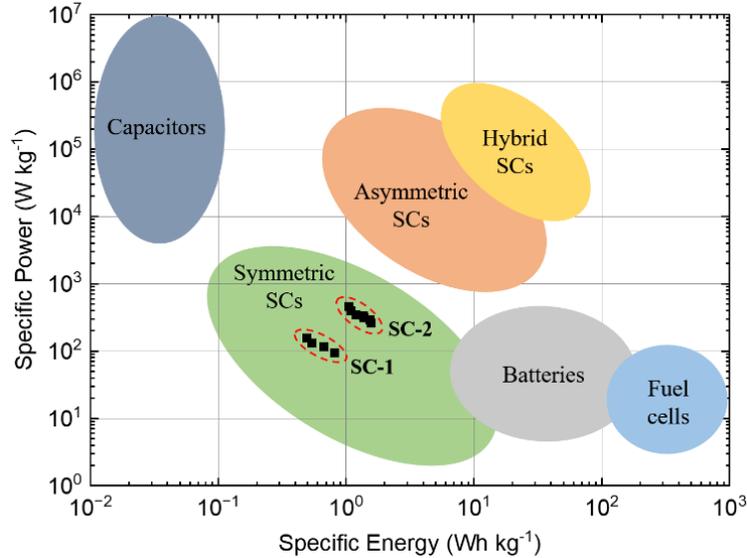

**Figure 7.** Ragone plot comparing the specific energy and specific power of the fabricated symmetric supercapacitors with other energy storage devices [2]. The positions of SC-1 and SC-2 are shown within the typical operating region of symmetric supercapacitors. The different data points for each device correspond to measurements performed at various current densities (A g$^{-1}$).

## 4. Conclusions

This study demonstrated the feasibility of fabricating fully sustainable symmetric supercapacitors using cotton textile waste as a renewable source for hydrogel electrolytes and chitosan-based biopolymer electrodes. The results confirmed that ionic modification of the cellulose hydrogel with ammonium thiocyanate ($NH_4SCN$) significantly enhances its structural integrity, ionic conductivity, and electrochemical response. The ion-modified device exhibited a reduced equivalent series resistance and a short time constant ($\tau \approx 3.2$ s), values comparable to those of commercial supercapacitors. After 1000 charge-discharge cycles, the ion-modified supercapacitor maintained stable operation with a 12.3 % increase in specific capacitance, confirming good electrochemical stability and efficient ion transport across the electrode-electrolyte interface. Spectroscopic and impedance analyses indicated that the moderate rise in resistance during cycling mainly arises from transient Fe–SCN coordination processes occurring when the liquid fraction expelled from the compressed hydrogel reaches the stainless-steel collector through the porous carbon electrode, rather than from degradation of the capacitive matrix. The overall capacitive behavior remained consistent, highlighting the mechanical and electrochemical resilience of the bio-based system. These results establish a strong proof of concept for integrating recycled cellulose hydrogels and chitosan-based carbon electrodes into efficient, metal-free, and environmentally responsible supercapacitors. The proposed design offers a viable pathway toward the development of high-performance, low-impact electrochemical energy-storage devices derived predominantly from renewable and bio-based materials.

**Conflicts of interest**

The authors declare that there are no conflicts of interest associated with this manuscript.

**Acknowledgments**

Luis T. Quispe gratefully acknowledges the support received for this work through the "Programa de Incorporación de Investigadores para fortalecer las áreas y líneas de investigación" of the "Instituto de Investigación Científica (IDIC)" at the Universidad de Lima. This research was also funded by the "Instituto de Investigación Científica" (IDIC) of the "Universidad de Lima", under project contract No. PI560052024.